\documentclass[prl,a4paper,twocolumn,superscriptaddress]{revtex4}
\usepackage{graphicx}
\usepackage{dcolumn}
\usepackage{bm}

\begin{document}

\title{Near-field photoluminescence study of InAs/AlGaAs quantum-dot-based
nanoclusters: band filling effect}
\author{Young-Jun Yu$^{1,2}$}
\author{Han-Eol Noh$^{1,2}$}
\author{Mun-Heon Hong$^{1,2}$}
\author{I. T. Jeong$^{2}$}
\author{J. C. Woo$^{2}$}
\author{Yeonsang Park$^{2}$}
\author{Heonsu Jeon$^{2}$}
\author{Wonho Jhe$^{1,2}$}
\email[Corresponding author:]{whjhe@snu.ac.kr}
\affiliation{$^{1}$Center for Near-field Atom-photon Technology and
$^{2}$School of Physics, Seoul National University, Seoul 151-742,
Korea}

\begin{abstract}
We have performed near-field spectroscopy and microscopy of the
InAs/AlGaAs quantum-dot-based nanoclusters. It is observed that
the photoluminescence spectra of spatially confined excitons in
the nanoclusters is blue-shifted up to 20~meV as the power
density is increased. In particular, the near-field
photoluminescence images have shown that excitons became
spatially confined gradually from lower energy state (1.4150~eV)
to higher energy state (1.4392~eV) as the excitation power is
increased, which is indicative of the band-filling effect of
semiconductor nanostructures.
\end{abstract}



\maketitle

The potential applications of semiconductor nanostructures to
optical nano-devices have been actively investigated. In
particular, self-assembly growth of semiconductor quantum dots
(QDs) has been the research focus as a method to fabricate optical
nano-devices. In order to integrate semiconductor QDs into the
nano-devices, their optical properties should be characterized on
various conditions. Among the optical properties of semiconductor
QDs, the band-filling
effect~\cite{Castrillo95,Sugisaki01,Masumoto02} is an interesting
phenomenon, in particular, in low-dimensional systems for which
the filling efficiency is enhanced. In this case, all the
available states of the lower exciton energy are fully occupied
under strong optical pumping. Thus a large number of confined
excitons are pumped to the higher states and then relax
radiatively to the ground state.

Previous works~\cite{Castrillo95,Wang01,Sugisaki01} on semiconductor
QDs have reported an interesting blueshift of the emission spectra
due to the band-filling effect as the excitation power is increased.
However, it is difficult to achieve the high spatial as well as
spectral resolution in optical images of the semiconductor
nanostructures by using the conventional optical characterizing
techniques.~\cite{Cingolani93,Xin99,Wang01} Therefore it is
necessary to employ near-field scanning optical microscopy (NSOM)
for the spatially resolved optical measurement of semiconductor
nanostructures.~\cite{pirson98,toda96,Saiki99,Yu03}

In this Letter, we report on the near-field photoluminescence (PL)
spectroscopy and microscopy of the InAs/AlGaAs QD-based
nanoclusters of several hundred nm in size by using
NSOM.~\cite{Eah02,Yu03} In particular, we have observed the
blueshift of the PL spectra as well as the gradual increase of
the dominant PL energies with the excitation power, which
manifests the band-filling effect of InAs/AlGaAs QD-based
nanoclusters.


\begin{figure}[b]
\begin{center}
\scalebox{0.48}{\includegraphics{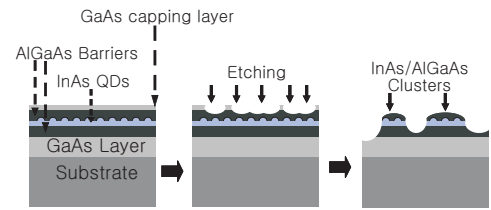}} \caption{Several hundred
nanometer-sized clusters that contain InAs/AlGaAs QDs, fabricated
by dry etching method. } \label{fig1}
\end{center}
\end{figure}

The InAs/AlGaAs QD-based nanoclusters were fabricated as follows.
First, the InAs/AlGaAs QD structure was grown on a 100~nm thick
GaAs buffer layer prepared on a GaAs (100) substrate as shown in
Fig.~1. The InAs QDs were embedded in a 50~nm thick AlGaAs
cladding layer. An additional 5~nm thick GaAs capping layer was
also grown by molecular beam epitaxy. The density of the InAs QDs
is typically about $5.5\times
10^{10}\textrm{cm}^{-2}$.~\cite{Kim00} Then, by dry etching of
the QD structure by using an inductively coupled plasma reactive
ion etcher (RIE)~\cite{Cho04}, InAs/AlGaAs QD structure was
etched down to the boundary region between the InAs QDs and the
lower AlGaAs barrier. Note that during the etching process, the
etched surface is not usually smooth in the nano-scale, which
results in the nanoclusters containing isolated InAs/AlGaAs QDs
as shown in Fig.~1.

For optical pumping of QDs, a Ti:Sapphire laser was used, which
was operated at the photon energy of 1.67~eV. This laser light was
coupled to a single-mode optical fiber and guided to a chemically
etched sharp fiber tip, where an 100-nm gold-coated aperture was
fabricated.~\cite{Ohtsu98} Such a nanometer-scale light source
generated by the aperture-fabricated fiber probe makes it possible
to excite nanoclusters in small area with nanometer-scale
position selectivity. The resulting PL was collected by the same
fiber so that any loss of spatial resolution due to diffusion
could be minimized.~\cite{Saiki99, Yu03} Both the sample and the
fiber probe were enclosed in a cryostat and kept at 77~K. The PL
was dispersed by a monochromator with 0.3~meV spectral resolution
and detected by a liquid-nitrogen-cooled charge-coupled device
camera to achieve high signal-to-noise ratio.

\begin{figure}[t]
\begin{center}
\scalebox{0.45}{\includegraphics{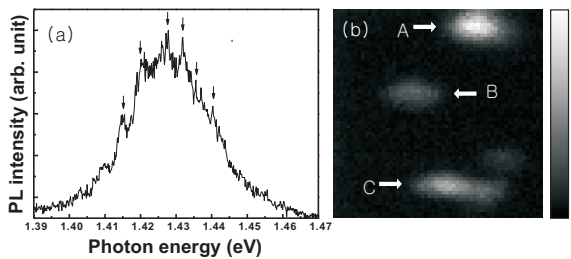}} \caption{(a)
Time-integrated PL spectra with collection time of 10~s at 77~K.
(b) Two-dimensional spatial images of near-field PL, obtained at
1.4271~eV PL energy. The scan area is $6~\mu\textrm{m}\times6~
\mu\textrm{m}$.} \label{fig2}
\end{center}
\end{figure}

Figure~2(a) shows the PL spectrum of an InAs/AlGaAs QD-based
nanocluster by employing NSOM. Although the PL peaks of single
InAs QDs (marked by arrows in Fig.~2(a)) are superimposed on the
broad background emission spectrum, they are not clearly resolved
due to inhomogeneous broadening at relatively high temperature
(77~K). Note that in order to perform high-resolution single QD
spectroscopy at this temperature, both an apertured probe with a
few hundred nm in diameter and a sample with a masking layer
having a few hundred nm apertures are necessary, which reduce the
number of excited single QDs as well as the transmission of the
excitation laser, as reported in Ref.~\onlinecite{Yu03}. The
spatial PL image of the nanoclusters was also obtained and
presented in Fig.~2(b), where the selected PL energy is 1.4271~eV
and the scan area is $6~\mu\textrm{m}\times6~ \mu\textrm{m}$. The
full width at half maximum (FWHM) along the horizontal axis of
the bright objects marked by the arrows A, B, and C in Fig.~2(b)
are 941, 1135 and 1471~nm, respectively. It indicates that the
nanoclusters were formed nonuniformly during the etching process.

\begin{figure}[b]
\begin{center}
\scalebox{0.45}{\includegraphics{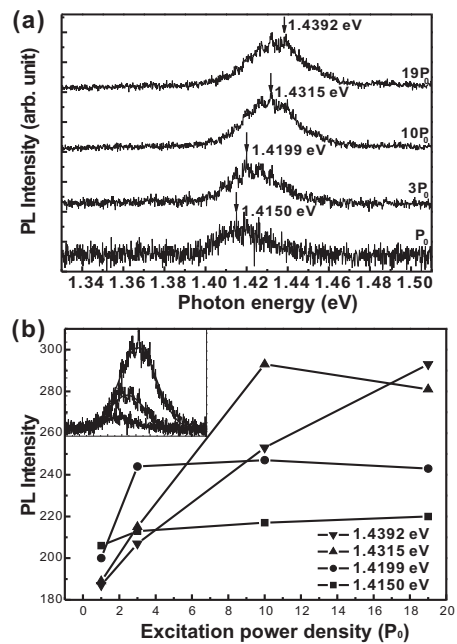}} \caption{(a)
Time-integrated PL spectra of an InAs/AlGaAs QD-based nanocluster,
collected for 1 s at 77~K, as the excitation power density is
increased from $P_{0}$ to $19 P_{0}$. (b) The excitation
power-density dependence of the intensity of the PL spectra of the
QD-based nanocluster. The inset shows the unnormalized PL spectra
of the nanocluster as the excitation power density is increased
from $P_{0}$, to $10 P_{0}$.} \label{fig3}
\end{center}
\end{figure}

Figure~3(a) presents the normalized PL spectra of a QD nanocluster
obtained by increasing the excitation power density from $P_{0}$ to
$19 P_{0}$, where $P_{0}$ is 8.6~kW/cm$^{2}$. As can be observed,
the blueshift ($\sim 17$~meV) and the linewidth broadening ($\sim
6$~meV) are obtained as the excitation power is increased. For
example, the center energy of the PL spectrum is shifted from
1.4175~eV at $P_{0}$ to 1.4335~eV at $19 P_{0}$. The inset of Fig.
3(b) shows the unnormalized PL spectra of QD nanocluster as the
excitation is increased from $P_{0}$ to $10 P_{0}$. Note that one
can clearly observe the plateau region of the PL spectrum ($\geq$
1.4315 eV) becomes increasingly broadened and also the PL intensity
is enhanced with the excitation power density. Figure 3(b) shows the
intensity saturation for the lower PL energy states (1.4150 and
1.4199~eV) as well as the intensity increase for the higher states
(1.4315 and 1.4392~eV) with the power density from $P_{0}$ to $19
P_{0}$. Note that these phenomena are well known characteristics of
the band-filling effect, as reported by other
works.~\cite{Castrillo95,Wang01,Sugisaki01,Masumoto02,Cingolani93,Xin99}
Note also that a possible temperature increase of QD-nanocluster due
to the heated probe at high excitation level may result in redshift
of the PL spectra~\cite{Jiang00}, which may eventually limit the
blueshift of PL spectrum.

\begin{figure}[t]
\begin{center}
\scalebox{0.45}{\includegraphics{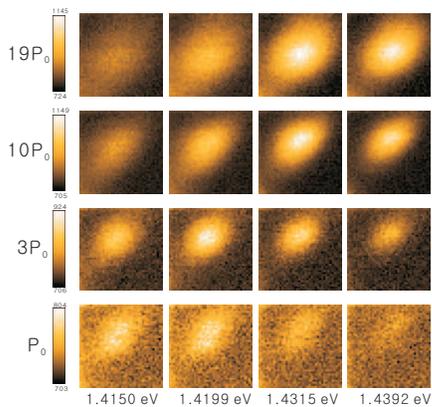}} \caption{Two-dimensional
near-field PL images obtained at different excitation levels and
PL energies of 1.4150, 1.4199, 1.4315 and 1.4392~eV, as
respectively marked by the arrows in Fig.~3(a). The scan area is
$2~\mu\textrm{m}\times2~ \mu\textrm{m}$. } \label{fig4}
\end{center}
\end{figure}
Figure~4 shows the two-dimensional spatial PL images, obtained for
several energy states (marked by the arrows in Fig.~3(a)) of the
nanocluster with respect to the excitation power density. The
near-field PL images were obtained at 25 nm interval in
$2~\mu\textrm{m}\times2~ \mu\textrm{m}$ scan area by using the
100-nm apertured fiber probe.~\cite{Eah02, Yu03} Note that, in
Fig.~4, the intensities of the PL images are displayed on the
same scale for each excitation level from $P_{0}$ to $19 P_{0}$.
The average FWHM along the horizontal-axis of the PL images is
about 600~nm. As can be observed, excitons become spatially
confined gradually from lower energy state (1.4150~eV) to higher
energy state (1.4392~eV) as the excitation power is increased from
$P_{0}$ to $19 P_{0}$, which may be attributed to the
band-filling effect of semiconductor nanostructures. In order to
have a quantitative understanding of the various PL images
presented in Fig.~4, we present the analyzed excitation-level
dependence of the intensity as well as the FWHM of the PL images
in Fig.~5.

\begin{figure}[t]
\begin{center}
\scalebox{0.5}{\includegraphics{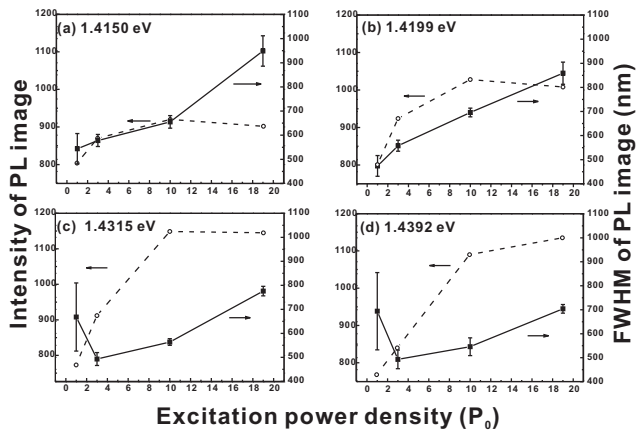}} \caption{Excitation power
density dependence of the intensity and the FWHM of the spatial PL
images shown in Fig.~4. } \label{fig5}
\end{center}
\end{figure}

Figures~5(a) and (b) show that the intensity of the PL images is
saturated as the excitation power density is increased, whereas the
FWHM is gradually increased. This indicates that the PL images at
1.4150 and 1.4199 eV energy in Fig.~4 are dispersed and spread with
the excitation power density. In case of Figs. 5(c) and (d), on the
other hand, the intensity increases more rapidly than in Figs. 5(a)
and (b). Especially, the PL intensity at 1.4392~eV in Fig. 5(d)
increases almost monotonically, in contrast to the height decrease
or saturation for other energy states in Figs. 5(a), (b), and (c),
at a given excitation level of $19 P_{0}$. Moreover, the FWHM and
its measurement uncertainty in Figs.~5(c) and (d) decrease rather
rapidly at the excitation power density from $P_{0}$ to $3P_{0}$,
but the FWHM starts increasing above $3P_{0}$ excitation. These
results indicate that while the PL images at 1.4315 and 1.4392~eV
energies in Fig.~4 appear at the excitation level from $P_{0}$ to
$3P_{0}$, the confining ratio of the excitons in these energy states
(1.4315 and 1.4392~eV) is enhanced beyond $3P_{0}$. In other words,
Fig.~4 shows that the dominant PL energies from the InAs/AlGaAs
QD-based nanocluster are gradually changed from the lower energy
states to the higher energy ones with the excitation pumping level.

In conclusion, we have studied the band-filling effect in the
InAs/AlGaAs QD-based nanoclusters, as observed by modification of
the spatial distributions of the PL images. In particular, the PL
energy states of the nanoculster are gradually changed from the
lower energy states to the higher energy states at a proper
excitation pumping level. The present work may be useful in
application of tunable energy states to optical nano-devices.

\begin{acknowledgments}
This work was supported by the Korean Ministry of Science and
Technology through the Creative Research Initiatives Program.
\end{acknowledgments}

\end{document}